\documentclass[aps,prc,twocolumn,showkeys,superscriptaddress]{revtex4-2}

\usepackage[colorlinks,allcolors={blue}]{hyperref}
\usepackage{natbib}
\usepackage{subfigure}
\usepackage{graphicx}
\usepackage{CJK}
\usepackage{url}
\usepackage{array}
\usepackage{makecell}
\usepackage{amsmath}
\usepackage{hyperref}

\begin{document}

\title{Nuclear mass predictions based on deep neural network and finite-range droplet model (2012)}

\newcommand{\ahku}{   \affiliation{Department of Physics, The University of Hong Kong, Pokfulam, 999077, Hong Kong}}
\newcommand{\autokyo}{  \affiliation{Department of Physics, Graduate School of Science, The University of Tokyo, Tokyo 113-0033, Japan}}
\newcommand{\aRIKEN}{  \affiliation{Interdisciplinary Theoretical and Mathematical Sciences Program (iTHEMS), RIKEN, Wako 351-0198, Japan}}

\author{To Chung Yiu}     \ahku
\author{Haozhao Liang}    \autokyo \aRIKEN
\author{Jenny Lee}        \ahku

\date{\today}

\begin{abstract}
A neural network with two hidden layers is developed for nuclear mass prediction, based on the finite-range droplet model (FRDM12). Different hyperparameters, including the number of hidden units, the choice of activation functions, the initializers, and the learning rates, are adjusted explicitly and systematically. 
The resulting mass predictions are achieved by averaging the predictions given by several different sets of hyperparameters with different regularizers and seed numbers. It can provide us not only the average values of mass predictions but also reliable estimations in the mass prediction uncertainties.
The overall root-mean-square deviations of nuclear mass have been reduced from $0.603$~MeV for the FRDM12 model to $0.200$~MeV and $0.232$~MeV for the training set and validation set, respectively.
\end{abstract}

\keywords{nuclear mass, machine learning, deep neural network}

\maketitle

\section{Introduction}

\label{Introduction}

Nuclear mass is one of the most fundamental nuclear properties~\cite{Lunney2003, Huang2021}.
It not only represents the static properties of nuclei, but also determines the reaction energies in different nuclear processes, such as $\beta$ decay, neutron capture, and fission~\cite{Mumpower2016}.
All these processes play important roles in the origin of elements and the abundance of elements in the universe~\cite{Burbidge1957, Mumpower2016, Martin2016}.

Nowadays, there are around $2500$ nuclei with  experimental masses. These nuclei are estimated to be only $27.8\%$ of around 9000 bounded nuclei estimated theoretically ~\cite{Moeller2016, Huang2021}. In order to have a better understanding of nuclear mass for all nuclei, theoretical mass models are required. There are several different theoretical models for mass prediction. Some of them are the microscopic models such as Hartree–Fock–Bogoliubov (HFB) mass models~\cite{Goriely2009, Goriely2013} and relativistic mean-field (RMF) mass models~\cite{Zhao2010}, while the others are the macroscopic-microscopic models such as finite-range droplet model~\cite{Moeller2016} and Weizsäcker–Skyrme (WS) model~\cite{Wang2014}. These theoretical models give a root-mean-square (RMS) deviation between the theoretical and experimental masses from $0.3$ to $2.3$~MeV~\cite{Niu2018}. To reduce RMS deviation, which is critical for a better description of final element abundance~\cite{Burbidge1957, Mumpower2016}, machine learning model is developed.

Neural network, one of the algorithms in machine learning, has been widely used in different research fields~\cite{Pederson2022, Kulik2022, Boehnlein2022}. From several recent studies, it has been proven that neural network is able to improve the accuracy of the models for several different nuclear properties, such as the $\beta$-decay half-lives~\cite{Niu2019,Minato2022}, neutron capture rate~\cite{Ma2019}, nuclear charge radii~\cite{Wu2020}, and ground-state and excited energies~\cite{Lasseri2020}. Neural network has also been used in nuclear mass prediction in several previous studies~\cite{Utama2017, Niu2018, Yueksel2021,Niu2022}. 
In particular, a significant improvement in mass predictions has been achieved by using several neural networks~\cite{Niu2022}.

Meanwhile, each neural network in these studies includes several kinds of hyperparameters, e.g., the number of hidden units, the choice of activation functions, the initializers, and the learning rates~\cite{Wu2021, Minato2022}.
These hyperparameters play very important roles in both the training process and the final performance of the neural network.
In other words, it is essential to investigate such hyperparameters in an explicit and systematic way. 

In this paper, a deep neural network will be used to build a mass model improving the current finite-range droplet model~\cite{Moeller2016}. Different hyperparameters will be in particular investigated in this study to achieve a better neural network model. Also, several different sets of hyperparameters will be used together to predict the nuclear masses and provide the uncertainty of the result.

The details about the neural network used in this study will be given in Sec.~\ref{Neural network model}. The results of the nuclear mass neural network model and its performance will be discussed in Sec.~\ref{Results}. Finally, a summary will be presented in Sec.~\ref{Summary}. For all the hyperparameters adjusted in this work, they will be given in Appendices~\ref{Activation function},~\ref{Learning rate},~\ref{Number of hidden units}, and~\ref{Initializers}.
\section{Neural network model} 

\label{Neural network model} 

\begin{figure}
    \centering
    \includegraphics[width=\linewidth]{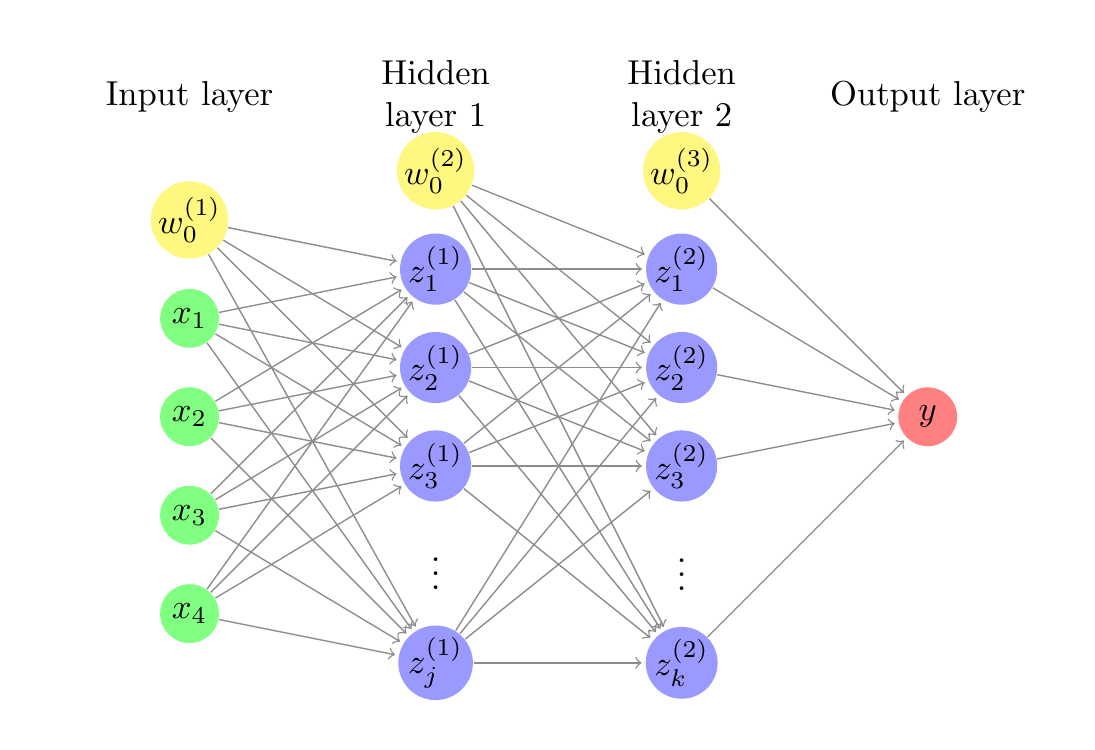}
    \caption{A deep neural network with four inputs, two hidden layers, and one output. The numbers of hidden units are $j$ and $k$ for the first and second hidden layers, respectively. For each of the input and hidden layers, there is a bias term shown with the yellow circle.}
    \label{NN}
\end{figure}

In this study, a deep neural network (DNN) consisting of an input layer, two hidden layers, and an output layer is used, as shown in Fig.~\ref{NN}. 
The relation between input, hidden, and output layers are as follows: 
\begin{align}
    \label{eq:DNN.1}
    z^{(1)}_j&=S(\sum_i w^{(1)}_{ji}x_i+w^{(1)}_{j0}), \\
    \label{eq:DNN.2}
    z^{(2)}_k&=\sigma(\sum_j w^{(2)}_{kj}z^{(1)}_j+w^{(2)}_{k0}), \\
    \label{eq:DNN.3}
    y_l&=\sum_l w^{(3)}_{lk}z^{(2)}_k+w^{(3)}_{l0},
\end{align} 
where $x_i$, $z^{(1)}_j$, $z^{(2)}_k$, and $y_l$ are the inputs, the hidden units in the first hidden layer, the hidden units in the second hidden layer, and the output of the model, respectively. 
Here, $\boldsymbol{w}=\{w^{(1)}_{ji},w^{(2)}_{kj},w^{(3)}_{lk},w^{(1)}_{j0},w^{(2)}_{k0}, w^{(3)}_{l0}\}$ are the parameters of the model, $S$ is the sigmoid function~\cite{TensorFlow:activations}  
\begin{equation}
    \label{eq:sigmoid}
    S(x)=\frac{1}{1+e^{-x}},
\end{equation}
and $\sigma$ is the softmax function~\cite{TensorFlow:activations}
\begin{equation}
    \label{eq:softmax}
    \sigma(x_k)=\frac{e^{x_k}}{\sum_k e^{x_k}},
\end{equation}
which makes the sum of $z^{(2)}_k$ in the second hidden equal to $1$. Hyperbolic tangent~\cite{TensorFlow:activations} has also been tried as the activation function for the model, 
\begin{equation}
    \label{eq:tanh}
    \tanh(x)=\frac{e^{x}-e^{-x}}{e^{x}+e^{-x}}.
\end{equation}
See the key reasons for choosing sigmoid and softmax functions as the activation functions in Appendix~\ref{Activation function}.

The inputs of the neural network are given by $\boldsymbol{x}=\{Z,N,\delta,P\}$, i.e., including neutron number $N$ and proton number $Z$, together with the information about nuclear pairing $\delta$ and shell effect $P$. In particular, the information about nuclear pairing is characterized by
\begin{equation}
    \label{eq: nuclear pairing}
    \delta=\frac{(-1)^N+(-1)^Z}{2},
\end{equation}
so $\delta=1$ for even-even nuclei, $\delta=0$ for odd-$A$ nuclei, and $\delta=-1$ for odd-odd nuclei.
The information about shell effect is characterized by
\begin{equation}
    \label{eq:shell effect}
    P=\frac{v_pv_n}{v_p+v_n},
\end{equation}
with $v_p,v_n$ being the difference between $Z$, $N$ and the closest magic number ($2$, $8$, $20$, $28$, $50$, $82$, $126$, or $184$), respectively~\cite{Kirson2008}. 

In this study, the finite-range droplet model (FRDM12) is used as the theoretical mass model~\cite{Moeller2016}. FRDM12 is a macroscopic–microscopic model, which includes the finite-range liquid-drop model in Eq.~\eqref{eq:frdm12.1} as the macroscopic model, 
\begin{align}
    \label{eq:frdm12.1}
    E_{\rm mac}(A,Z)=&~a_vA+a_sA^{2/3}+a_3A^{1/3}B_k+a_0A^0 \nonumber \\ &+E_c-c_2Z^2A^{1/3}B_r-c_4\frac{Z^{4/3}}{A^{1/3}}\nonumber \\ &-c_5Z^2\frac{B_wB_s}{B_1}+f_0\frac{Z^2}{A}-c_a(Z-N)\nonumber \\&+W+E_{\rm{pairing}}-a_{el}Z^{2.39},
\end{align}
and the folded-Yukawa single-particle potential as the microscopic corrections~\cite{Moeller2016}. FRDM12 has a relatively small root-mean-square (RMS) deviation between the experimental and theoretical mass when compared with other theoretical mass models. Its overall RMS deviation is $0.603$~MeV, while the RMS deviation of nuclear mass for RMF model is $2.269$~MeV~\cite{Niu2018}. Besides, the FRDM model has been used to develop different models for $\beta$-decay properties such as the FRDM+QRPA model~\cite{Mumpower2016}. Therefore, the improved FRDM12 with the neural network model can be used to improve those $\beta$-decay models.

The target we aim to achieve in this work is the mass residuals between the theoretical FRDM12 mass $M_{l,\rm{FRDM12}}$~\cite{Moeller2016} and the experimental mass $M_{l,\rm{exp}}$~\cite{Huang2021}, 
\begin{equation}
    \label{eq:massdifference}
    t_l=M_{l,\rm{FRDM12}} - M_{l,\rm{exp}}.
\end{equation}
To achieve the target mass residuals from the neural network, different hyperparameters of the model are required to be adjusted. The mean-square-error,
\begin{equation}
    \label{eq:mse}
\rm mse(\boldsymbol{x},\boldsymbol{w})= \sum_l(t_l-y_l)^2,
\end{equation}
is used as the loss function, with $y_l$ and $t_l$ obtained from Eqs.~\eqref{eq:DNN.3} and~\eqref{eq:massdifference}, respectively. Adam algorithm~\cite{TensorFlow:Adam} with a learning rate of 0.01 is used to adjust the parameters $\boldsymbol{w}$ to achieve a smaller loss function. The corresponding key reasons are shown in Appendix~\ref{Learning rate}. For the hidden layers, each of the hidden layers consists of 22 hidden units and a bias, see the key reasons in Appendix~\ref{Number of hidden units}. Therefore, there are $639$ parameters in $\boldsymbol{w}$ in total for the model.

For the initial parameter $\boldsymbol{w}$ before the training, there are three different initializers tried in this study. They include the standard normal initializer, Glorot normal initializer, and zeros initializer~\cite{TensorFlow:initializers}. The standard normal initializer distributes the initial parameters from a standard normal distribution with the mean being $0$ and the standard deviation being $1$. The Glorot normal initializer distributes the initial parameters from a normal distribution with the mean being $0$ and the standard deviation equaling to $\sqrt{{2}/(f_{\rm{in}}+f_{\rm{out}})}$. In the above equation, $f_{\rm in}$ and $f_{\rm out}$ are the numbers of input units and output units of that layer, respectively~\cite{TensorFlow:initializers}. The zeros initializer sets the initial parameters equal to zero~\cite{TensorFlow:initializers}.
In this study, the initial parameters $\boldsymbol{w}$ before the training are generated from the standard normal distribution for $\{w^{(1)}_{ji}, w^{(2)}_{kj}\}$, and Glorot normal distribution for $\{w^{(1)}_{j0}, w^{(2)}_{k0}\}$, see key reasons in Appendix~\ref{Initializers}.  

The experimental data of this work are from the atomic mass evaluation of 2020 (AME2020)~\cite{Huang2021}. There are in total $2457$ nuclei with $Z,N \geq 8$. They are separated into a training set and a validation set randomly. $1966$ nuclei ($80\%$) are randomly selected for the training set, while the remaining $491$ nuclei ($20\%$) are in the validation set. 

To achieve a neural network with less variance in the prediction of mass, $33$ sets of hyperparameters with different regularizers and seed numbers are used for the training. There are three different types of regularizers used in this study. The first type is without using any regularizer. The second type is the L2 regularizer. It includes an additional term in the loss function~\cite{TensorFlow:regularizers},
\begin{equation}
    \label{L2 regularizer}
    L_{\rm{L2}}(\boldsymbol{w})=\lambda \times \sum(w)^2,
\end{equation}
with $\lambda$ being a hyperparameter controlling the rate of regularization. The third type is the orthogonal regularizer. It will encourage the basis of the output space of that layer to be orthogonal to each other~\cite{TensorFlow:regularizers}. The hyperparameter $\lambda$ is used to control the rate of regularization.

In the following calculations, $20000$ epochs are run for each training and the smallest RMS deviation between $t_l$ and $y_l$ of the validation set in different epochs is used to determine the performance of the set of hyperparameters. Only those sets of hyperparameters giving an RMS deviation of the validation set smaller than $0.228$~MeV (reduction of RMS deviation over $60\%$ compared with the FRDM12 prediction) are selected. The average prediction from the selected sets of hyperparameters is the model output $y$. Summing the model output y and the theoretical FRDM12 mass $M_{k,\rm{FRDM12}}$ will give us the model mass prediction for the nuclei. 
\section{Results and discussion}

\label{Results} 

\begin{figure*}
    \centering
    \includegraphics[width=0.9\linewidth]{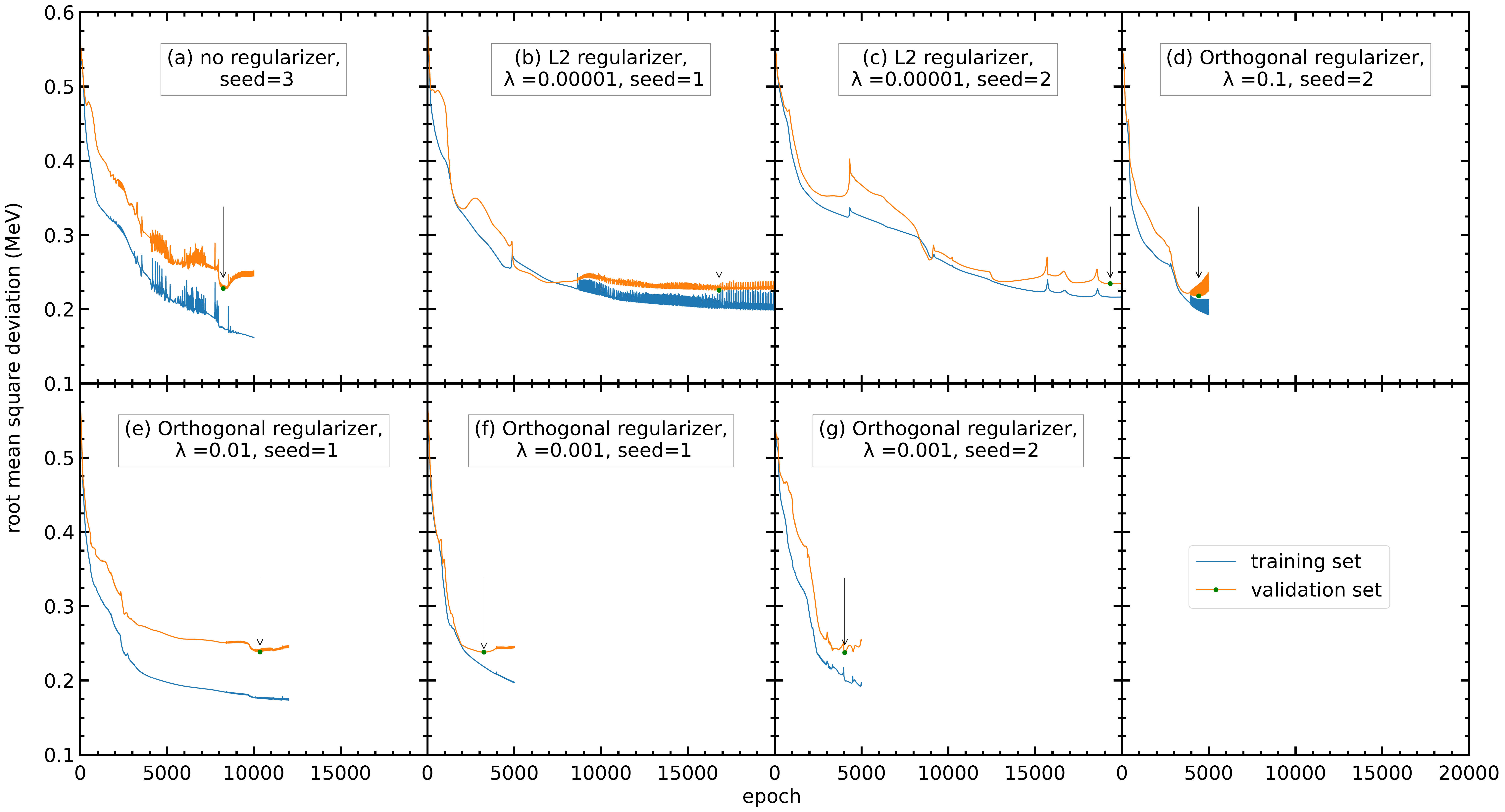}
    \caption{The RMS deviations of the training and validation sets with different sets of hyperparameters. Only the seven sets of hyperparameters with a reduction in RMS deviation larger than $60\%$ are selected as follows: (a) No regularizer, seed $=3$, (b) L2 regularizer, $\lambda=0.00001$, seed $=1$, (c) L2 regularizer, $\lambda=0.00001$, seed $=2$, (d) Orthogonal regularizer, $\lambda =0.1$, seed $=2$, (e) Orthogonal regularizer, $\lambda=0.01$, seed $=1$, (f) Orthogonal regularizer, $\lambda=0.001$, seed $=1$, and (g) Orthogonal regularizer, $\lambda=0.001$, seed $=2$.
    The blue and orange curves represent the RMS deviations of the training and validation sets, respectively. The arrow points to the lowest RMS deviation of the validation set within the $20000$ epochs.}
    \label{Epoch}
\end{figure*}

Of the $33$ sets of hyperparameters, seven sets of them are with a reduction in RMS deviation larger than $60\%$, as shown in Fig.~\ref{Epoch}. We can see that different regularizers will speed up or slow down the training of the model, compared to the model without a regularizer in panel (a). Moreover, since the seed number will affect the random numbers generated during training and the initial values of the parameters, different seed numbers are tried for the model. From Table~\ref{table:epoch}, the RMS derivations have average reductions from $0.603$~MeV to $0.200$~MeV and $0.232$~MeV, having improvements of $66.8\%$ and $61.5\%$ for the training and validation sets, respectively. It shows that the neural network can improve the accuracy of the FRDM12 model after adjusting the hyperparameters.

\begin{table*}
    \begin{center}
    \caption{The RMS deviations of nuclear mass between the experimental data from AME2020 \cite{Huang2021} and the model predictions. The original RMS deviation between experimental data and FRDM12 \cite{Moeller2016} is $0.603$~MeV. For each set of hyperparameters, the RMS deviations for the training and validation sets are denoted by $\sigma_{\rm training}$ and $\sigma_{\rm validation}$, respectively. The last two columns show the reduction of RMS deviations from the original deviation $0.603$~MeV to $\sigma_{\rm training}$ and $\sigma_{\rm validation}$, respectively. }
        \begin{tabular}{c c c c c} 
         \hline
         Hyperparameters & $\sigma_{\rm training}$~(MeV) & $\sigma_{\rm validation}$~(MeV) & $\Delta\sigma_{\rm training} $ ($\%$) & $\Delta\sigma_{\rm validation} $ ($\%$) \\ [0.5ex] 
         \hline\hline
         \makecell{No regularizer,\\ seed $=3$} & $0.175$ & $0.228$ & $71.0$ & $62.2$ \\ 
         \hline
         \makecell{L2 regularizer,\\ $\lambda=0.00001$, seed $=1$} & $0.204$ & $0.226$ & $66.2$ & $62.5$  \\
         \hline
         \makecell{L2 regularizer,\\ $\lambda=0.00001$, seed $=2$} &  $0.216$ & $0.235$ & $64.2$ & $61.0$ \\ 
         \hline
         \makecell{Orthogonal regularizer,\\ $\lambda=0.1$,  seed $=2$} & $0.212$ & $0.218$ & $64.8$ & $63.8$ \\
         \hline
         \makecell{Orthogonal regularizer,\\ $\lambda=0.01$, seed $=1$} & $0.176$ & $0.238$ & $70.8$ & $60.5$ \\
         \hline
         \makecell{Orthogonal regularizer,\\ $\lambda=0.001$, seed $=1$} & $0.219$ & $0.238$ & $63.7$ & $60.5$ \\
         \hline
         \makecell{Orthogonal regularizer,\\ $\lambda=0.001$, seed $=2$} & $0.201$ & $0.238$ & $66.7$ & $60.5$ \\
         \hline 
         \makecell{Average of the above sets} & $0.200$ & $0.232$ & $66.8$ & $61.5$ \\[1ex]
         \hline
        \end{tabular}
    \label{table:epoch}
    \end{center}
\end{table*}

\begin{figure*}
    \centering
    \includegraphics[width=0.9\linewidth]{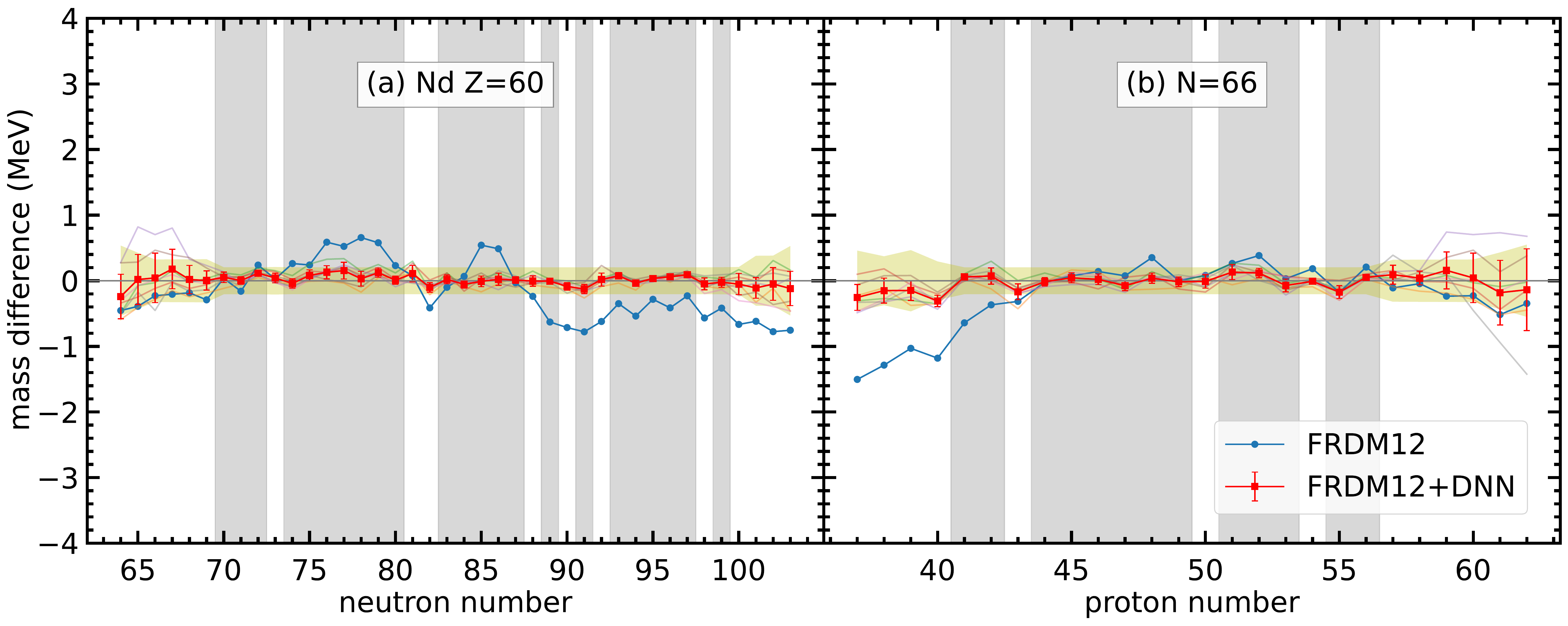}
    \caption{Mass difference between the DNN mass data and the experimental data, $t_k=M_{k,DNN} - M_{k,exp}$, for (a) Nd isotopes and (b) $N=66$ isotones. The blue curves represent the mass difference between the FRDM12 model and the experimental data. 
    The semi-transparent curves in the background represent the mass difference between the FRDM12+DNN mass model and the experimental data with the seven sets of hyperparameters, respectively. 
    The red curves with uncertainties represent the mass difference for the average of the FRDM12+DNN mass model. The error bars are calculated from the standard deviation of the mass difference from the $7$ sets of hyperparameters.
    The gray-hatched regions represent the training set. The yellow-hatched regions represent the mass uncertainties by including the average RMS deviations of seven sets of hyperparameters together with the experimental uncertainties.}
    \label{Nd&N66}
\end{figure*}

For most of the isotopes and isotones, the neural network model can give a good performance in mass prediction. For example, from Fig.~\ref{Nd&N66}, the average of the selected seven sets of hyperparameters has a much better overall performance in the mass difference compared to the FRDM12 model for the Nd isotopes and $N=66$ isotones. The data in the training set is represented in the gray-hatched regions in Fig.~\ref{Nd&N66}. It is shown that the mass prediction in the training set is, in general, better than that in the validation set. For the regions far away from the training set, such as $N<68$ for Nd isotopes and $Z>59$ for $N=66$ isotones, the mass difference is still within the mass uncertainty as shown in the yellow-hatched regions, so the neural network model is still with a good performance in those regions. However, the error bar for the neural network model is larger in regions far away from the training set. It shows that the mass prediction for different sets of hyperparameters in those regions is with large variation. Therefore, multiple sets of hyperparameters are selected, and averaging the prediction of the seven sets of hyperparameters can provide us with the uncertainty of nuclear mass prediction for different nuclei. The larger the error bar for the nuclei, the less confidence we have in the mass prediction.

\begin{figure}
    \centering
    \includegraphics[width=0.9\linewidth]{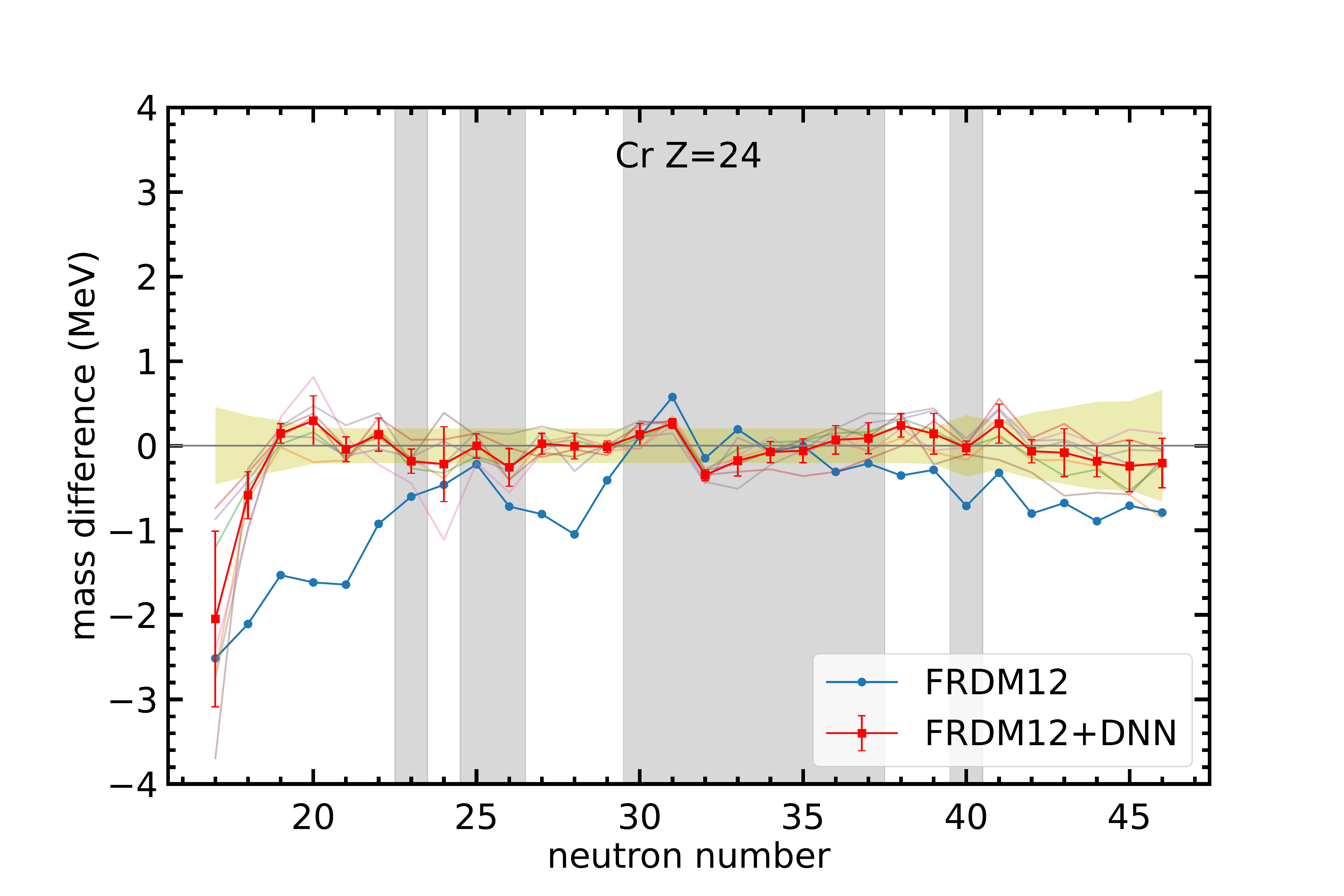}
    \caption{Same as Fig.~\ref{Nd&N66}, but for Cr isotopes.}
    \label{Cr}
\end{figure}

Furthermore, there are several nuclei in which the masses from the FRDM12 model are significantly different from the experimental data. For example, the mass difference can be up to $2.5$~MeV. It is interesting to see whether or not the neural network can significantly improve the mass predictions for those specific nuclei.

For Cr isotopes shown in Fig.~\ref{Cr}, FRDM12 cannot give a good mass prediction for the region $N<22$. In this region, the neural network model can still provide a better mass prediction compared with the FRDM12 model. However, the mass difference of the two nuclei $N=17$ and $18$ is not within the yellow-hatched region. Although the mass prediction for the two nuclei $N=17$ and $18$ is not as good as other nuclei, the neural network model can improve the mass difference from $-2.5$~MeV to $-2$~MeV and from $-2.1$~MeV to $-0.6$~MeV, respectively. It shows that the neural network model can improve mass prediction, even though the original theoretical model cannot provide an accurate prediction.

\begin{figure}
    \centering
    \includegraphics[width=0.9\linewidth]{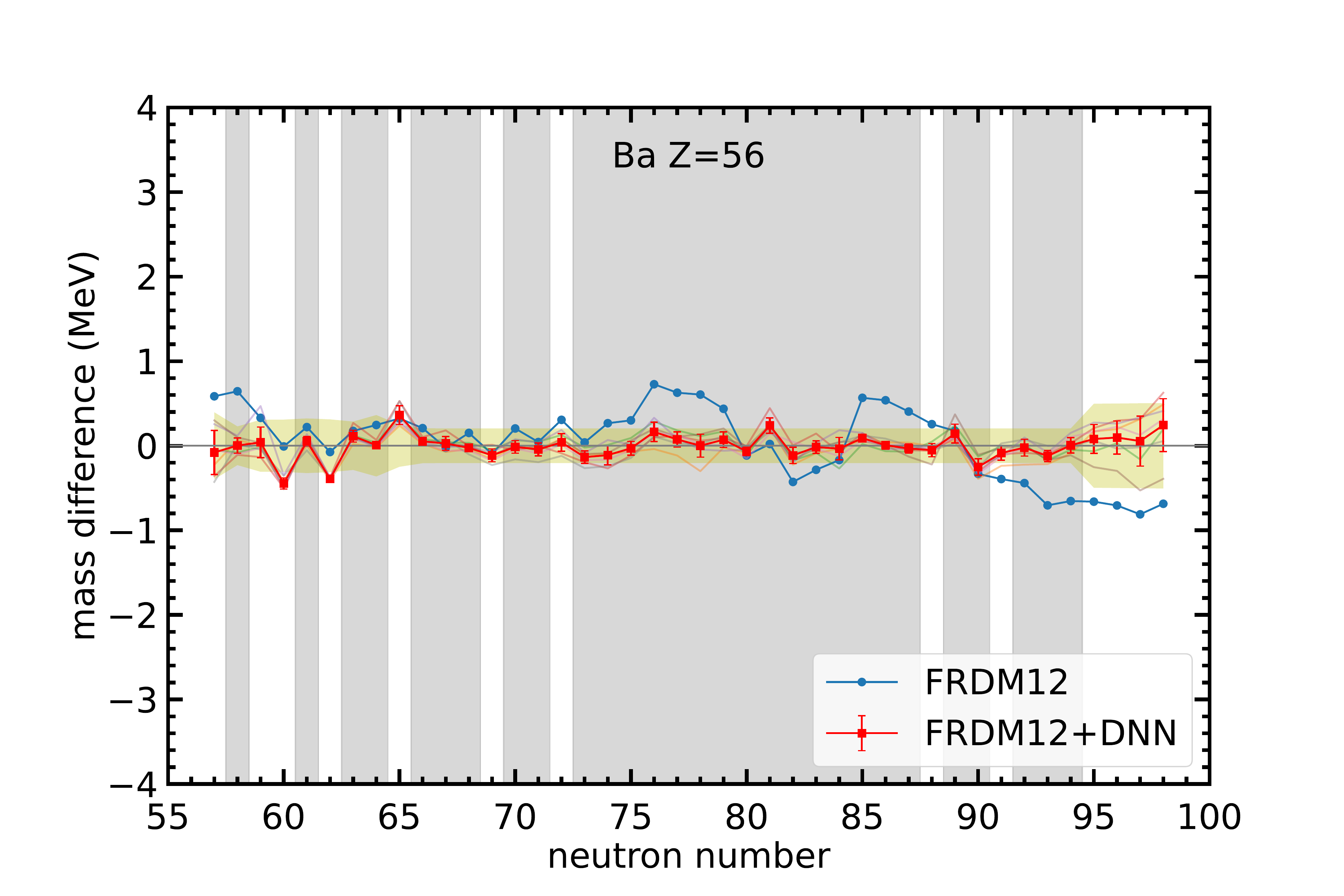}
    \caption{Same as Fig.~\ref{Nd&N66}, but for Ba isotopes.}
    \label{Ba}
\end{figure}

Finally, we also need to see if there are any nuclei in which the mass prediction becomes worse after applying the neural network model.
For Ba isotopes in Fig.~\ref{Ba}, it shows a good agreement in the mass prediction for the training set and large-neutron-number regions. However, for $N=60$ and $62$, the neural network gives worse predictions compared to the FRDM12 model. It may happen because of two reasons. First, the nuclei $N=60$ and $62$ are in the validation set, which are not involved in the training of the model. Therefore, the overall performance of the nuclei in the validation set is not as good as the nuclei in the training set. Second, this neural network model still cannot tackle the odd-even staggering in light nuclei. Therefore, these two even-even nuclei do not have a better mass prediction after using the neural network, while the nearby even-odd nuclei in the training set have a better performance.

\section{Summary and perspectives} 

\label{Summary} 

The deep neural network has been applied to study the nuclear mass based on the FRDM12 model. Different hyperparameters, including activation functions, learning rate, the number of hidden units, and the initializers, are adjusted in a systematic way to achieve a better performance of the model, as shown in Appendices~\ref{Activation function},~\ref{Learning rate},~\ref{Number of hidden units}, and~\ref{Initializers}. Finally, seven sets of hyperparameters with different regularizers and seed numbers have been selected.
It is important to find that averaging the predictions given by several different sets of hyperparameters can provide us not only the average values of mass predictions but also reliable estimations in the mass prediction uncertainties.

With the neural network, the RMS deviations between the experimental mass and theoretical mass have been reduced from $0.603$~MeV to $0.200$~MeV and $0.232$~MeV for the training set and validation set, respectively. For most of the nuclei, this DNN model can give a better mass prediction compared with the FRDM12 model. Even for those nuclei which have a poor mass prediction in the FRDM12 model such as $^{41}$Cr and $^{42}$Cr, the DNN model can still reduce the RMS deviation and achieve a better mass prediction. However, there are still several nuclei with worse mass prediction compared with the FRDM12 model. It may happen in the validation set with unsolved odd-even staggering problems. Further studies for those nuclei are required in the future to improve the model, such as a better description of the odd-even staggering in nuclei.

In the future, the same technique can be applied to other physics quantities related to nuclei. With the nuclear mass predicted in this study and adjusting different hyperparameters, DNN models for different physics quantities, such as $\beta$-decay half-life and $\beta$-delayed neutron emission probability, can be generated and their performance can be compared with other current theoretical models to investigate if this algorithm can improve the present predictions.

\begin{acknowledgments}
H.L. acknowledges the JSPS Grant-in-Aid for Early-Career Scientists under Grant No.~18K13549, the JSPS Grant-in-Aid for Scientific Research (S) under Grant No.~20H05648, and the RIKEN Pioneering Project: Evolution of Matter in the Universe.
J.L. acknowledges the Research Grants Council (RGC) of Hong Kong with a grant of General Research Funding (GRF-17312522).
\end{acknowledgments}

\appendix

\section{Performance with different activation function} 

\label{Activation function} 

In this appendix, we investigate explicitly the impact of the activation functions on the performance of the training.

\begin{figure}
    \centering
    \includegraphics[width=0.9\linewidth]{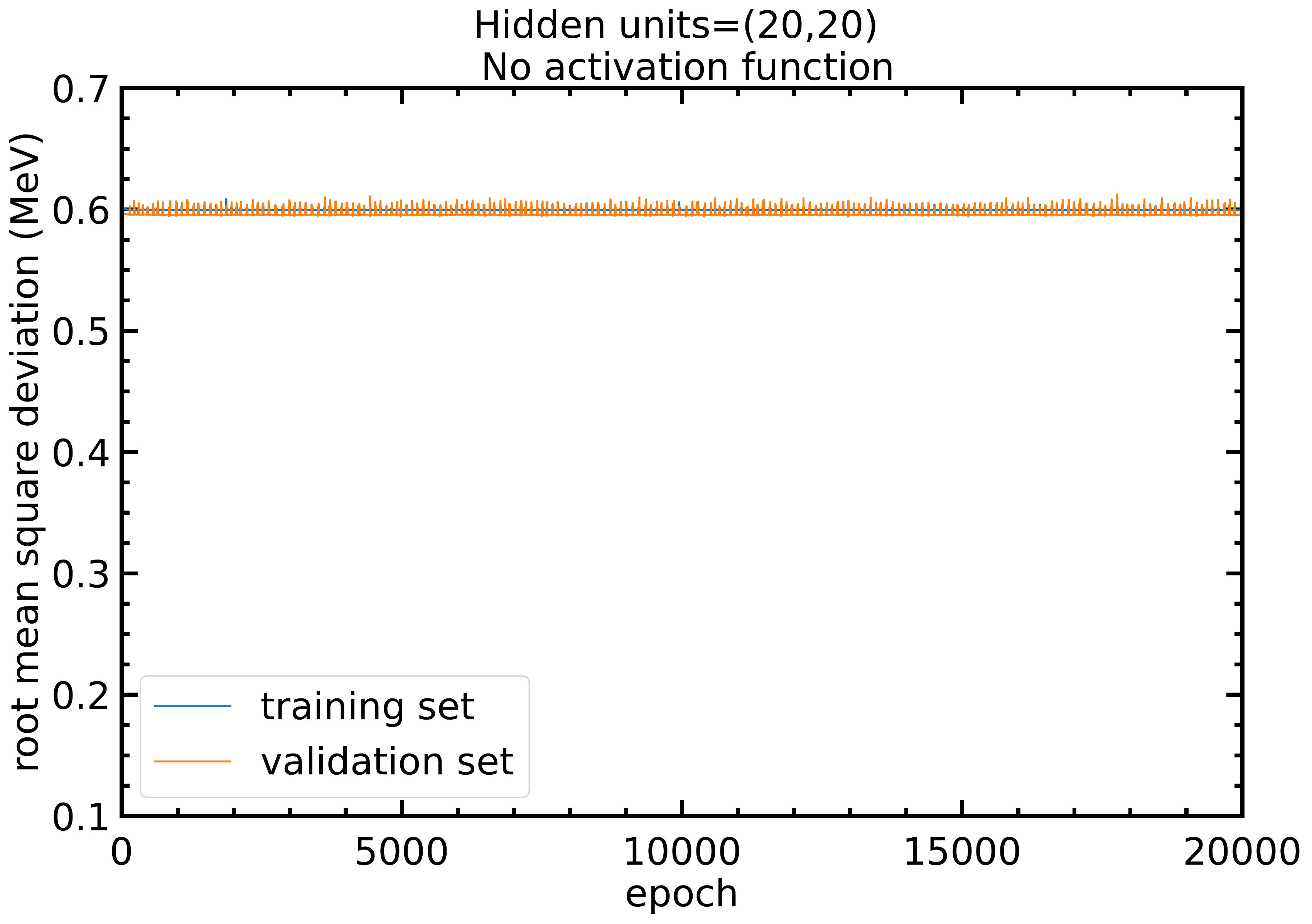}
    \caption{The RMS deviations of the training and validation sets in $20000$ epochs with $20$ hidden units in both hidden layers, and no activation function. The blue and orange curves represent the RMS deviations of the training and validation sets, respectively.}
    \label{Epoch_hidden20_act44}
\end{figure}

The neural network model is able to approximate non-linear relations between inputs and outputs because of the use of the activation functions. Without using an activation function, the model is just a linear regression that cannot approximate non-linear relations. Figure~\ref{Epoch_hidden20_act44} shows that, without the activation function, the RMS deviation of the model will remain at around $0.6$~MeV and no improvement can be made at all.
 
\begin{figure*}
    \centering
    \includegraphics[width=0.9\linewidth]{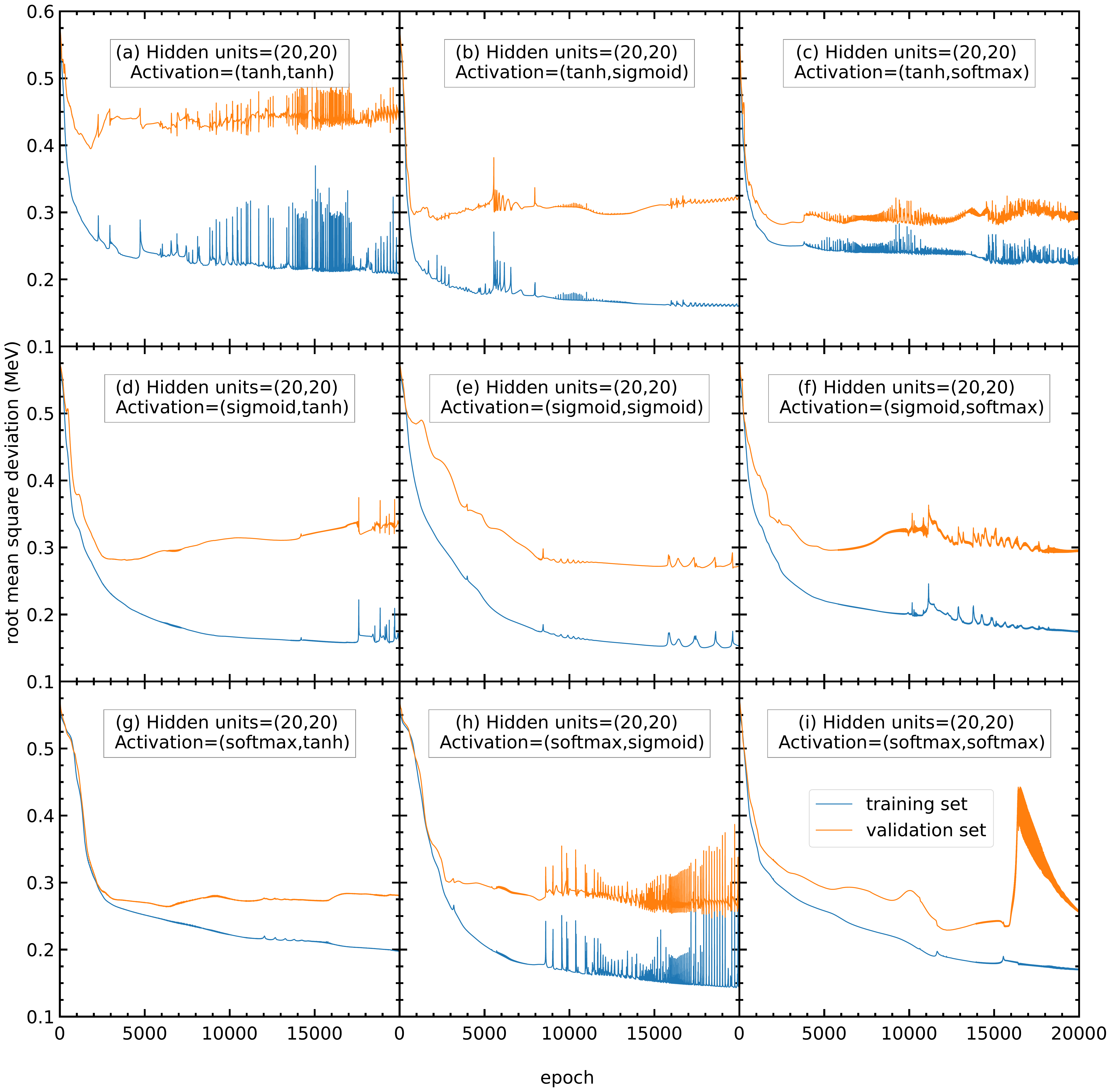}
    \caption{The RMS deviations of the training and validation sets in $20000$ epochs with $20$ hidden units in both hidden layers. 
    The activation functions for the two hidden layers are selected to be (a) (tanh, tanh), (b) (tanh, sigmoid), (c) (tanh, softmax), (d) (sigmoid, tanh), (e) (sigmoid, sigmoid), (f) (sigmoid, softmax), (g) (softmax, tanh), (h) (softmax, sigmoid), and (i) (softmax, softmax), respectively.
    The blue and orange curves represent the RMS deviations of the training and validation sets, respectively.}
    \label{Epoch_hidden20_act11_33}
\end{figure*}

\begin{figure*}
    \centering
    \includegraphics[width=0.9\linewidth]{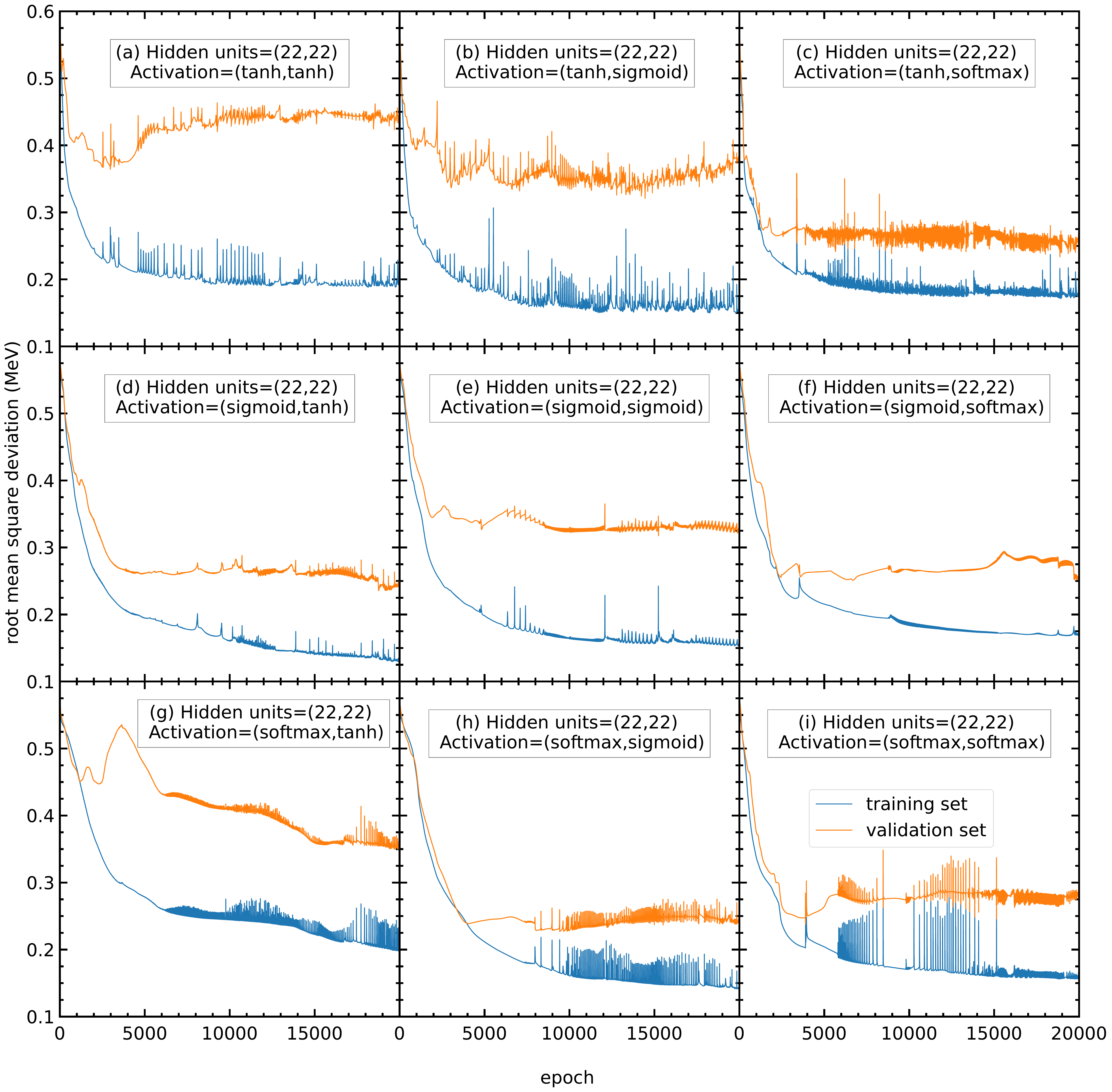}
    \caption{Same as Fig.~\ref{Epoch_hidden20_act11_33}, but the number of hidden units in both hidden layers is $22$.}
    \label{Epoch_hidden22_act11_33}
\end{figure*}

Therefore, models with different activation functions are trained in this study.
From Fig.~\ref{Epoch_hidden20_act11_33} with $20$ hidden units, the RMS deviations of panels (a), (c), (h), and (i) are highly oscillating, and thus the corresponding activation functions are not recommended. In contrast, panels (b), (d), (e), (f), and (g) show a steady RMS deviation without large oscillation. 

To find out the best activation functions for the present mass models, results from Fig.~\ref{Epoch_hidden22_act11_33} with $22$ hidden units are examined. Panel (b) is oscillating and not recommended. Panels (e) and (g) show RMS deviations larger than $0.3$~MeV, which is larger than the RMS deviations of panels (d) and (f), so panels (e) and (g) are not recommended. Panels (d) and (f) show small RMS derivations and less oscillation, so they are recommended for this mass model. In this study, the activation functions for the two hidden layers are chosen to be (f) sigmoid and softmax functions for all the following calculations, respectively. If there are any future studies, the activation functions can be chosen as (d) sigmoid and tanh functions, respectively.

\section{Performance with different learning rates} 

\label{Learning rate} 

In this appendix, we investigate explicitly the impact of the learning rate in the algorithm of the model on the performance of the training.

\begin{figure*}
    \centering
    \includegraphics[width=0.9\linewidth]{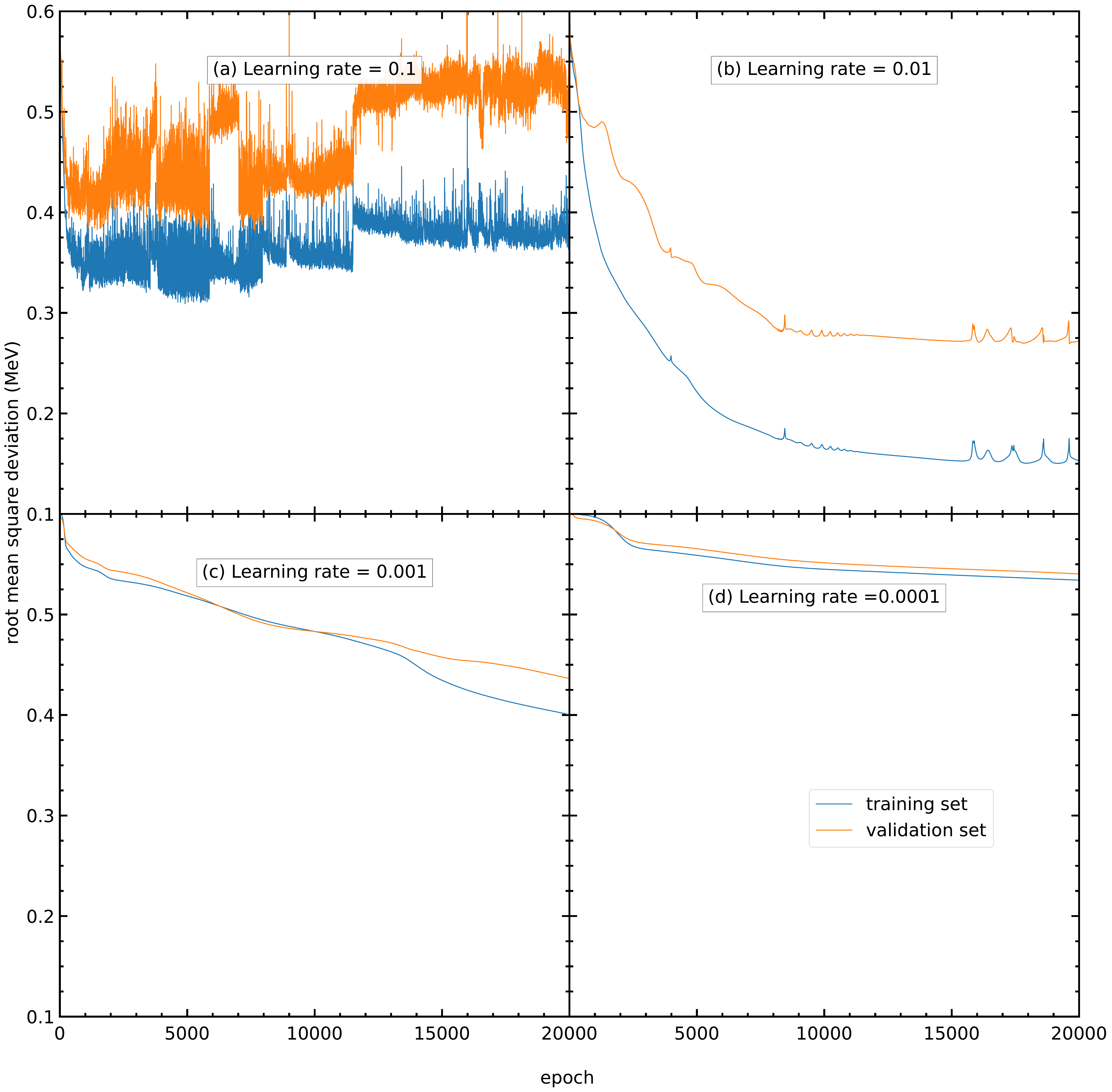}
    \caption{The RMS deviations of the training and validation sets in $20000$ epochs using Adam algorithm as the optimizer with the learning rate (a) $0.1$, (b) $0.01$, (c) $0.001$, and (d) $0.0001$. The blue and orange curves represent the RMS deviations of the training and validation sets, respectively.}
    \label{Epoch_learningrate}
\end{figure*}

In Fig.~\ref{Epoch_learningrate}, the RMS deviations of the training and validation sets using the Adam algorithm with the learning rates $0.1$, $0.01$, $0.001$, and $0.0001$ are shown.
It is seen in Fig.~\ref{Epoch_learningrate}(a) that a model with a large learning rate makes the model parameters oscillate and is unable to reach the minimum, so its RMS deviations are highly oscillating with large values.
A model with a smaller learning rate makes the number of epochs used to reach the lowest RMS deviation larger, and it is more time-consuming. Using the learning rate $0.01$, the RMS deviation for the validation set reaches around $0.27$~MeV after $20000$ epochs as shown in Fig.~\ref{Epoch_learningrate}(b).
In contrast, using the learning rates $0.001$ and $0.0001$, the RMS deviations for the validation set are still around $0.43$ and $0.54$~MeV after $20000$ epochs as shown in Figs.~\ref{Epoch_learningrate}(c) and \ref{Epoch_learningrate}(d), respectively.
Therefore, the learning rate is selected as $0.01$ for all the following calculations in this study.
\section{Performance with different number of hidden units} 

\label{Number of hidden units} 

In this appendix, we investigate explicitly the impact of the number of hidden units on the performance of the training.

\begin{figure*}
    \centering
    \includegraphics[width=0.9\linewidth]{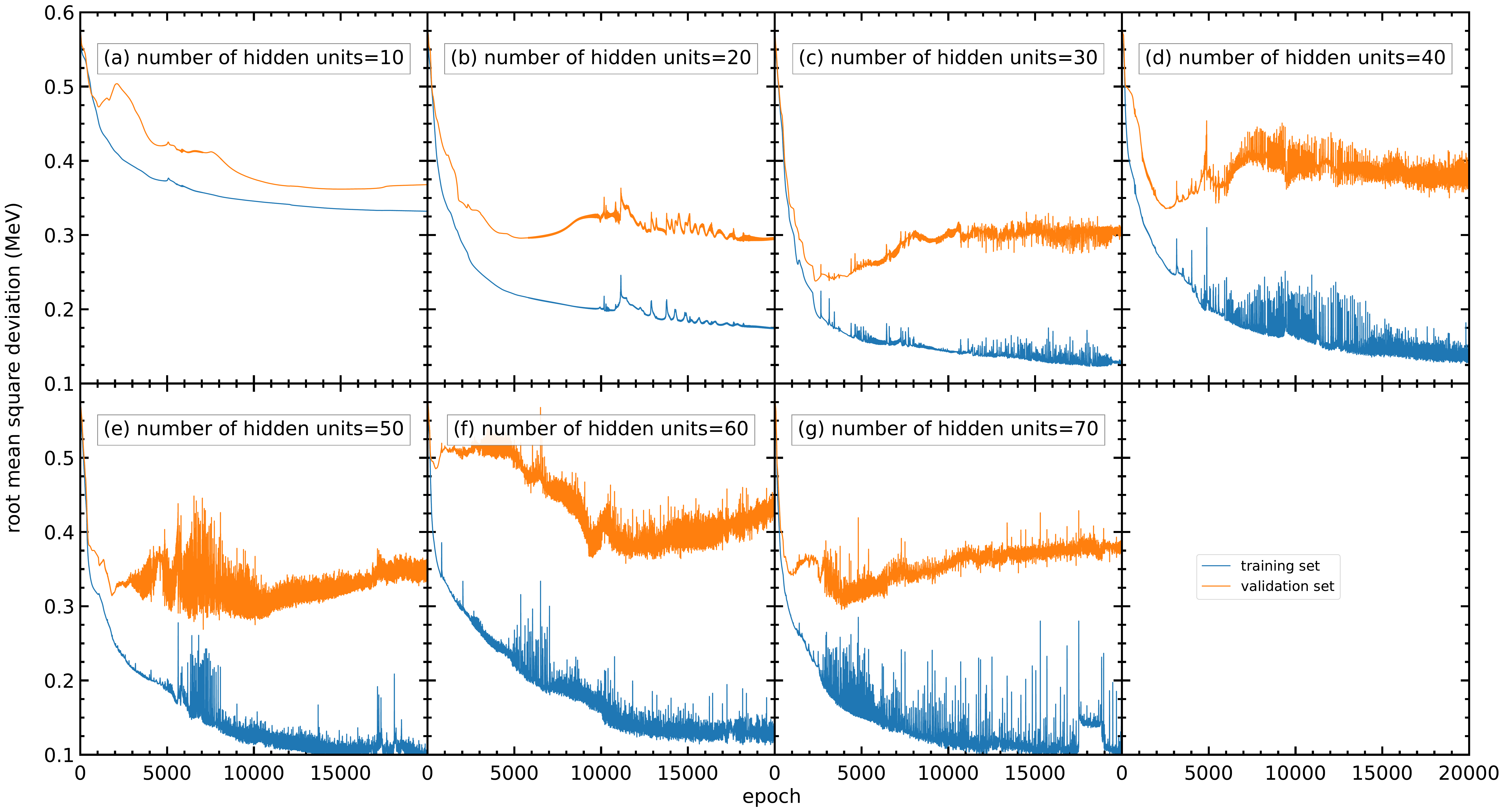}
    \caption{The RMS deviations of the training and validation sets in $20000$ epochs with different numbers of hidden units per hidden layer within $10$--$70$: (a) $10$, (b) $20$, (c) $30$, (d) $40$, (e) $50$, (f) $60$, and (g) $70$. The blue and orange curves represent the RMS deviations of the training and validation sets, respectively.}
    \label{Epoch_10_70}
\end{figure*}

In Fig.~\ref{Epoch_10_70}, the RMS deviations of the training and validation set with different numbers of hidden units per hidden layer are shown, where the numbers of hidden units per hidden layer are changed from $10$ to $70$.
Generally speaking, a model with a small number of hidden units may make the number of parameters not enough for the model and be unable to reach a small RMS deviation.
It is confirmed by Fig.~\ref{Epoch_10_70}(a) that the RMS deviation for the validation set for $10$ hidden units per hidden layer remains around $0.37$~MeV after $10000$ epochs.

For the model with a larger number of hidden units, the number of parameters will become larger, and the time used for training will become longer.
Moreover, since the data in the training set in general include several kinds of noise, the large number of hidden units may lead to the overfitting problem and cause a poor prediction outside the training set. 
As shown in panels (e), (f), and (g), although the RMS deviations for the training set are still decreasing after $10000$ epochs, the RMS deviations for the validation set are increasing in that region, resulting in a poor prediction of mass.

On the other hand, panel (b) shows less oscillation in the RMS deviation compared to panels (c) and (d).
Meanwhile, $30$ or more hidden units show no improvement in RMS deviation compared with $20$ hidden units.
For the above reasons, the number of hidden units around $20$ is selected.

\begin{figure*}
    \centering
    \includegraphics[width=0.9\linewidth]{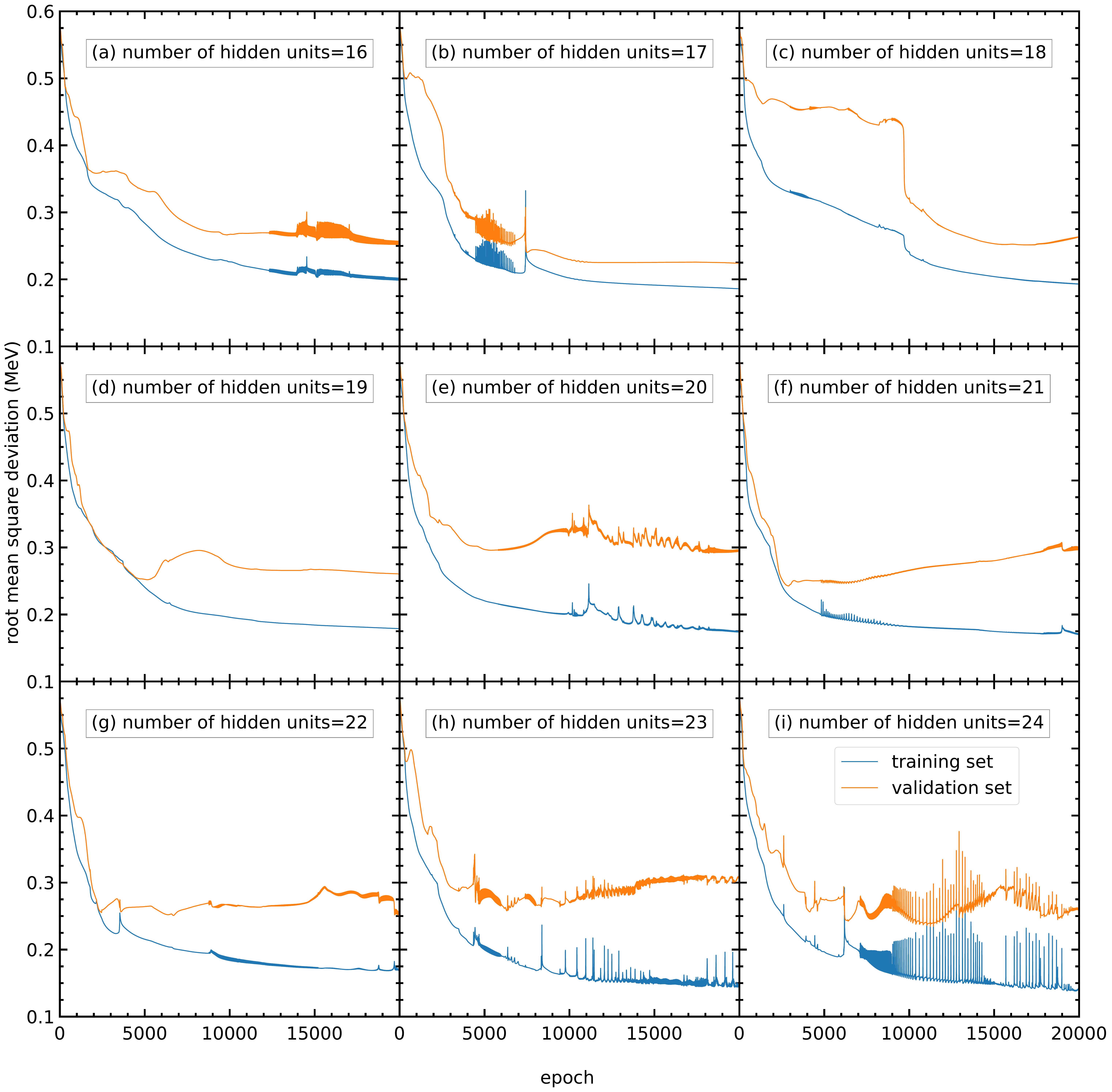}
    \caption{The RMS deviation of the training and validation set in $20000$ epochs with different numbers of hidden units per hidden layer within $16$--$24$: (a) $16$, (b) $17$, (c) $18$, (d) $19$, (e) $20$, (f) $21$, (g) $22$, (h) $23$, and (i) $24$. The blue and orange curves represent the RMS deviations of the training and validation sets, respectively.}
    \label{Epoch_16_24}
\end{figure*}

To find a more precise number of hidden units, models with different numbers of hidden units from $16$ to $24$ are trained and their performance is shown in Fig.~\ref{Epoch_16_24}.
It is seen that the performance of the models with different numbers of hidden units is similar, with around $0.25$~MeV in the RMS deviation of the validation set. However, it shows that, for models with $23$ and $24$ hidden units per hidden layer, the RMS deviations are oscillating, so the models with $23$ and $24$ hidden units per hidden layer are not recommended. For the remaining models with similar performance, the number of hidden units per hidden layer is chosen to be $22$, as it can provide more parameters without large oscillation or overfitting.
\section{Performance with different initializers} 

\label{Initializers} 

In this appendix, we investigate explicitly the impact of the initializers on the performance of the training.

\begin{figure*}
    \centering
    \includegraphics[width=0.9\linewidth]{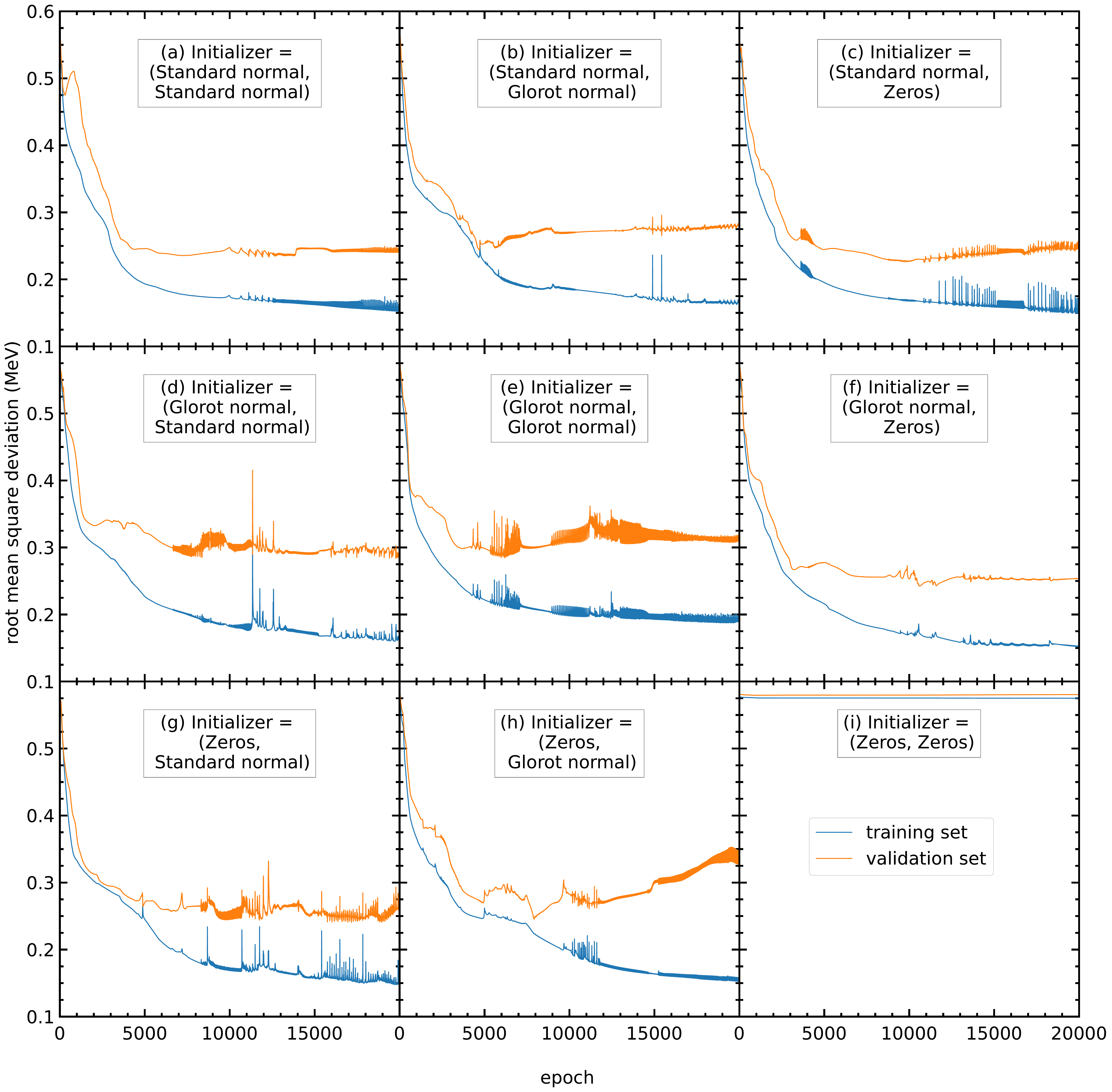}
    \caption{The RMS deviations of the training and validation sets in $20000$ epochs with different initializers for the parameters $\{w^{(1)}_{ji},w^{(2)}_{kj}\}$ and bias $\{w^{(1)}_{j0},w^{(2)}_{k0}\}$. The initializers adopted here include the standard normal, glorot normal, and zeros distributions \cite{TensorFlow:initializers}. The blue and orange curves represent the RMS deviations of the training and validation sets, respectively.}
    \label{Epoch_initial}
\end{figure*}

The initial values of the parameters also affect the performance of the model. 
For example, when all the parameters are initialized with the same value, all the hidden units in the same hidden layer will be identical to each other. Such kind of model is equivalent to a model with 1 hidden unit in each hidden layer. The small number of hidden units makes such a model with poor performance and cannot predict the output. 
As shown in Fig.~\ref{Epoch_initial}(i), the RMS deviations are remaining at around $0.575$~MeV during the training. 

Therefore, we train the models with different sets of initial parameters ($\{w^{(1)}_{ji}, w^{(2)}_{kj}\}$, $\{w^{(1)}_{j0}, w^{(2)}_{k0}\}$). For different sets of initial parameters trained, panels (a),(b), (c), (f), (g), and (h) give the RMS deviations of the validation set around $0.23$~MeV. Among these 6 sets of initial parameters, panels (a), (b), and (f) give a steady RMS deviation curve during the whole training. Therefore, these three sets of initial parameters are recommended. In this work, initial parameters from panel (b) are selected for all the following calculations. If there are any future studies, panels (a) and (f) can be used as the initial parameters.


%

\end{document}